\documentclass[vecphys]{svmult}

\usepackage{makeidx}         
\usepackage{graphicx}        
\usepackage{multicol}        
\usepackage[bottom]{footmisc}
\usepackage{amsmath}
\usepackage{bm}

\makeindex             

\begin{document}

\title*{Eigenmode of Decision-By-Majority Process on Complex Networks}
\author{Makoto Uchida\inst{1}\and
Susumu Shirayama\inst{2}}
\institute{ School of Engineering, the University of Tokyo. 5-1-5 Kashiwanoha, Kashiwa, Chiba 277-8568 Japan.  \texttt{uchida@race.u-tokyo.ac.jp}
\and Research into Artifacts, Center for Engineering (RACE), the University of Tokyo. 5-1-5 Kashiwanoha, Kashiwa, Chiba 277-8568 Japan. \texttt{sirayama@race.u-tokyo.ac.jp}}
%
%
\maketitle


The nature of dynamics of opinion formation modeled 
as a decision-by-majority process in complex networks
is investigated using eigenmode analysis.
Hamiltonian of the system is defined, and estimated
by eigenvectors of the adjacency matrix constructed 
from several network models.
The eigenmodes of initial and final state of 
the dynamics are analyzed by numerical studies. 
We show that the magnitude of the largest eigenvector 
at the initial states are key determinant for
the resulting dynamics. 

\section{Introduction}
\label{sec:1}

Many phenomena related to spreading, synchronization and 
collective dynamics have been studied from the viewpoint
of complex networks. These studies have revealed some
significant influences of network structure on the phenomena, 
and contributed to the development of complex network analysis
\cite{Boccaletti:2006}.
In such studies, a simple interaction model has been utilized.
This is because complex phenomena often emerges from simple
interactions.

We focus on dynamics of opinion formation in networks. 
One of the approaches for this dynamics is referred to
spin systems. In such systems, local interactions are
modeled using spin-like variables located at
the vertices of the networks. 
Complex dynamics have been explored both analytically and 
computationally. Strong dependency on properties of the 
networks has been pointed out 
\cite{Boyer:2003,Castellano:2005,Suchecki:2005,Sood:2005}. 
In some networks, the final consensus state does not
always have all the vertices with the same opinion 
\cite{Castellano:2005,Suchecki:2005}. 
Moreover, it has been found that history of convergence to
a steady state or a quasi-steady state depends on the topology 
of networks, as well as the local rule \cite{Lambiotte:2006}.
%
%
%
We also studied an effect of initial conditions on Glauber dynamics
for Ising model \cite{Uchida:2007}. We found that non-random initial
conditions play a key role to determine the final consensus state,
and that the final fraction of each opinion $r'$ is a function of
initial fraction of the corresponding opinion $r$. 
%

In this paper, we will manifest the mechanism that all
the vertices with the same opinion does not always appear 
in the final state, and consider the origin of the specific
$r-r'$ relations under different initial conditions.
We assume that the mechanism is related to the eigenmodes 
correspond to each eigenvectors of adjacency matrix of
networks.
Dynamics of opinion formation is analyzed using eigenmode analysis.

\section{Models}
\label{sec:2}

%
Let $v_{i}$ denote a vertex in a network.  For each vertex,
there are two possible states which represent two opposite opinions.
These states are represented by a spin-like variable:
$\sigma_{i}(t) = \pm{}1$ at a time $t$. 
$\sigma_{i}(t+1)$ is determined only by the values of neighboring 
vertices of $v_{i}$. We adopt a local rule driven by decision-by-majority
process. It is described by Eqn.~(\ref{eqn:majority})
\begin{equation}
\sigma_{i}(t+1)=\operatorname{sgn}\left\{\sum_{j=1}^{n}{a_{ij}\sigma_{j}(t)}\right\},
\label{eqn:majority}
\end{equation}
where $a_{ij}$ is a component of the adjacency matrix ${\bf A}$
of the network, which takes the value of $1$ if an edge exists 
between vertices $v_{i}$ and $v_{j}$, otherwise $0$.
$\sigma_{i}(t)$ is updated synchronously at each step $t$;
the values at all vertices are updated simultaneously 
as $t$ progresses.
%
%
%
%
In our previous studies \cite{Uchida:2006,Uchida:2007}, 
we found that depending on initial conditions and network structures,
the state of the system at $t=\infty{}$ from Eqn.~(\ref{eqn:majority}) 
becomes either a fully ordered state in which all vertices have 
the same state, or a metastable state in which two states co-exist. 
In this paper, we analyze the final states by eigenmode analysis. 

We assume that the system controlled by the local rule (Eqn.~(\ref{eqn:majority}))
evolves as the energy becomes lower, and the most stable state 
appears. 
First, we define a Hamiltonian $\mathcal{H}$ of the system as follows: 
\begin{equation}
\mathcal{H} = -\frac{J}{2}\sum_{i,j}\sigma_{i}\sigma_{j}a_{ij},
\label{eqn:hamiltonian}
\end{equation}
where $\sigma{}_{i}$ and $\sigma{}_{j}$ are the states of vertices $v_{i}$ and $v_{j}$, 
$J$ is a positive constant. This Hamiltonian is regarded as the energy of our system. 
Equation (\ref{eqn:hamiltonian}) means that the local energy becomes lower if vertices with
the same state are adjacent.
Secondly, we introduce another expression of $\mathcal{H}$ using vector form.
Let $\bm{s}$ be the spin state vector at time $t$;
$ \bm{s} = \begin{pmatrix}\sigma{}_{1}(t) & \sigma_{2}(t) 
& \cdots{} & \sigma_{n}(t) \end{pmatrix}^{T}$, 
where $n$ is the number of vertices.
%
%
%
%
%
%
The spin state vector $\bm{s}$ is expanded in the vector subspace 
spanned by eigenvectors of $\bm{A}$ as $\bm{s} = \sum_{i=1}^{n} c_{i} \bm{v}_i$,
%
%
where $\bm{v}_i$ is the $i$th eigenvector of $\bm{A}$, and 
$c_{i}$ is the $i$th coefficient corresponding to the eigenvector.
%
%
%
%
%
%
%
%
After some calculus, we obtain 
\begin{eqnarray}
\mathcal{H} = -\frac{J}{2} \sum_{i}c_{i}^{2}\lambda{}_{i},
\label{eqn:h_eig}
\end{eqnarray}
where $\lambda{}_{i}$ is the $i$th eigenvalue of $\bm{A}$.
%
%
For simplicity, we deal with undirected and unweighted networks,
in which $\bm{A}$ becomes a real symmetric matrix, 
and all the eigenvalues are real. 
Thus, the eigenvalues are labeled in descending order without loss of generality; 
$\lambda{}_{1} \geq \lambda{}_{2} \geq \cdots{} \geq \lambda{}_{n}$.
Considering that $\lambda{}_{1}$ is the largest eigenvalue, it can be 
expected that the most stable state is excited by such the $\bm{s}$ 
that gives the maximal $c_{1}$. 
%
%
It can be shown that the eigenmode corresponds to the largest eigenvalue dominantly
appears in the system. 
However, we cannot explain a metastable state (in which two states 
co-exist) by this eigenmode. 
%
%
In the following section, 
we numerically study the eigenmodes using several models of complex networks.

\section{Numerical Studies}

\subsection{Network Models and Initial Conditions}


In this paper the following network models are studied: 
Erd{\"o}s-Reny{\'i} random graph (ER), Watts-Strogatz (WS) model 
Barab{\'a}si-Albert (BA) model, Klemm-Eg{\'i}ulz (KE) model
and Connecting Nearest Neighbor (CNN) model.
The number of vertices is $n = 3000$, and $\langle{} k \rangle{} = 10$ for the average degree.
%
Each model has different structural characteristics.
%
See Ref. \cite{Boccaletti:2006} for the details. 
%
%
%
%
In such complex networks, vertices are not interconnected homogeneously.
%
We assume that the resultant dynamics depends on the distribution of 
initial state denoted by $\bm{s}_{0}$ owing to this heterogeneity. 
In order to study the dependency of $\bm{s}_{0}$, we consider arbitrary 
distributions of initial states 
according to several types of {\itshape centrality}; degree centrality 
and closeness centrality of the vertices, as well as random distribution.
See Ref.~\cite{Brandes:2001} for the detail of 
these centrality measures.
At $t=0$, fraction $r$ and distribution of the vertices
which $\sigma{}$ takes $+1$ are determined.
%
Corresponding to the centrality measures, the $rn$ vertices 
with the largest centrality are assigned as $\sigma{}(0) = +1$,
while the remaining $(1-r)n$ vertices are assigned
$\sigma{}(0) = -1$.

\subsection{ Relationships between Eigenmodes and Dynamics}  


\begin{figure}[htb]
\begin{center}
 \begin{tabular}{ccc}
   \resizebox{0.32\linewidth}{!}{ \includegraphics[]{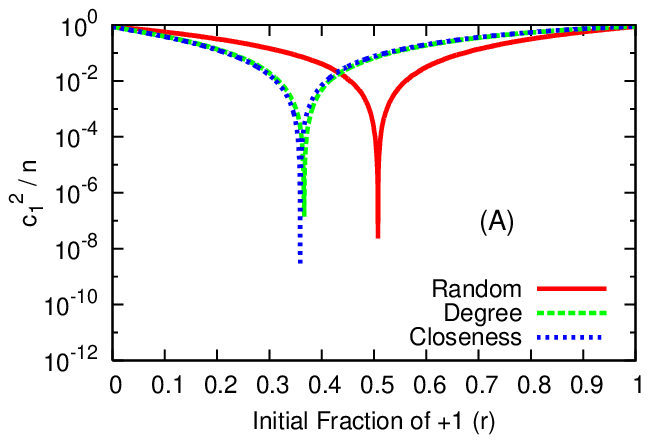}}  &
   \resizebox{0.32\linewidth}{!}{ \includegraphics[]{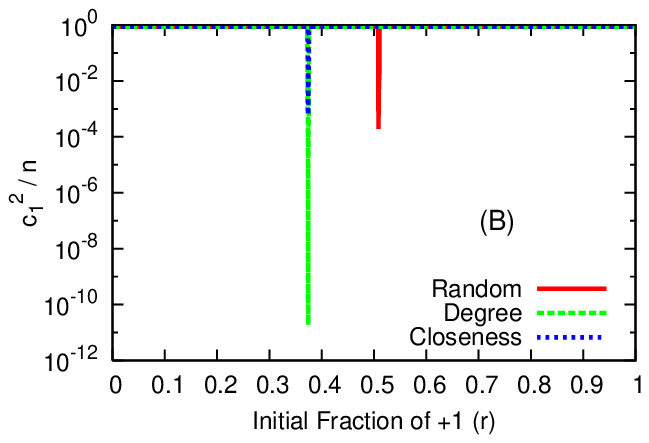}}  &
   \resizebox{0.32\linewidth}{!}{ \includegraphics[]{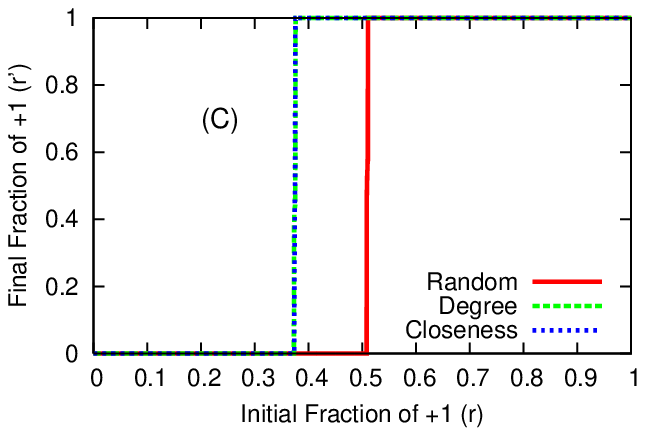}}  \\
   \multicolumn{3}{c}{ ER random graph } \\

   \resizebox{0.32\linewidth}{!}{ \includegraphics[]{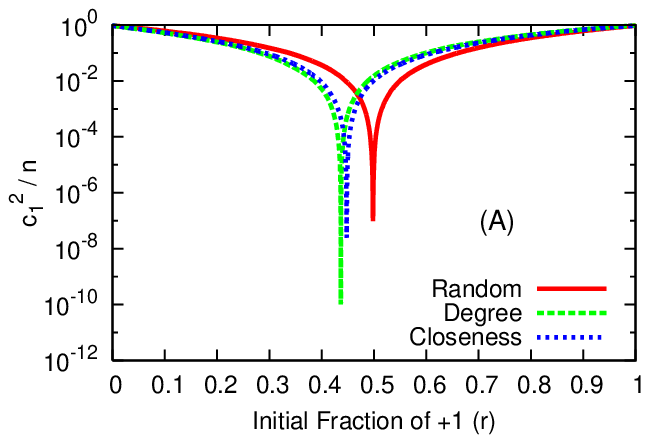}}  &
   \resizebox{0.32\linewidth}{!}{ \includegraphics[]{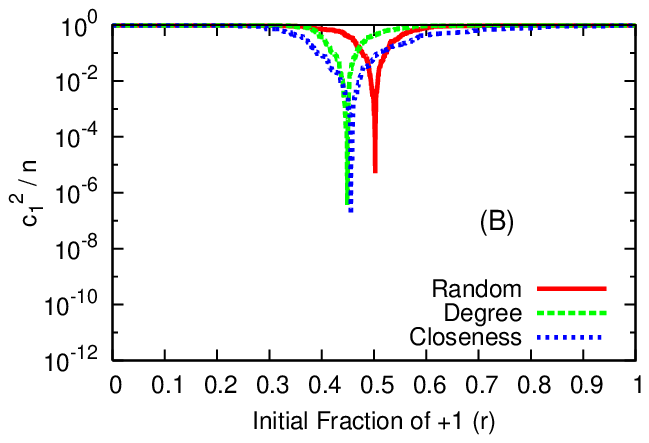}}  &
   \resizebox{0.32\linewidth}{!}{ \includegraphics[]{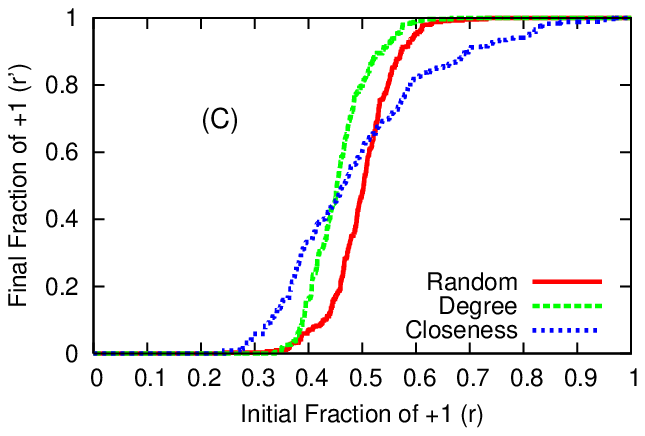}}  \\
   \multicolumn{3}{c}{ WS network } \\

   \resizebox{0.32\linewidth}{!}{ \includegraphics[]{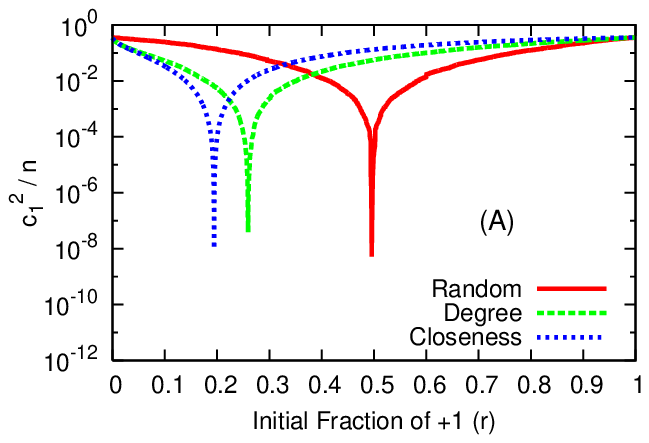}}  &
   \resizebox{0.32\linewidth}{!}{ \includegraphics[]{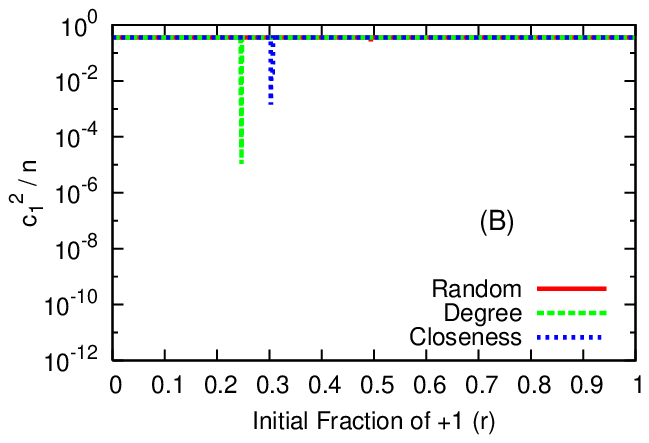}}  &
   \resizebox{0.32\linewidth}{!}{ \includegraphics[]{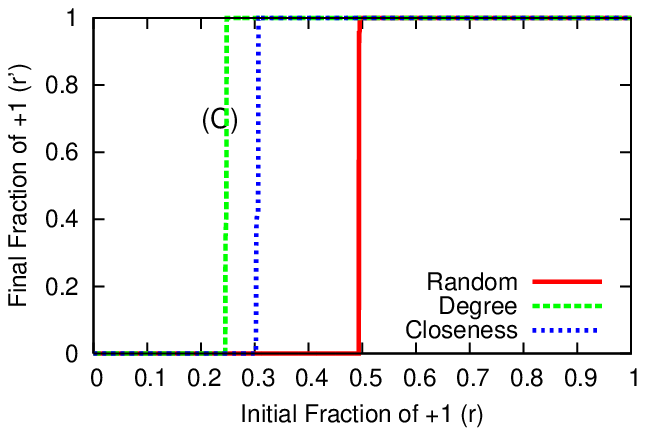}}  \\
   \multicolumn{3}{c}{ BA network } \\

   \resizebox{0.32\linewidth}{!}{ \includegraphics[]{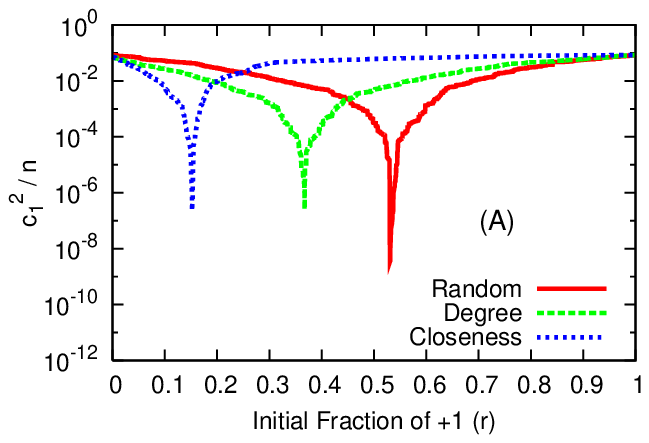}}  &
   \resizebox{0.32\linewidth}{!}{ \includegraphics[]{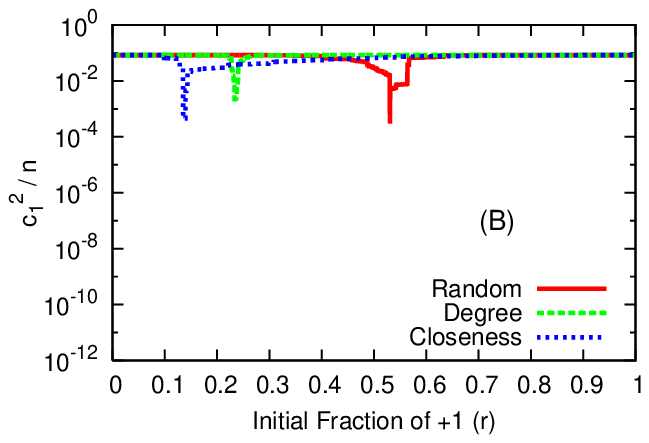}}  &
   \resizebox{0.32\linewidth}{!}{ \includegraphics[]{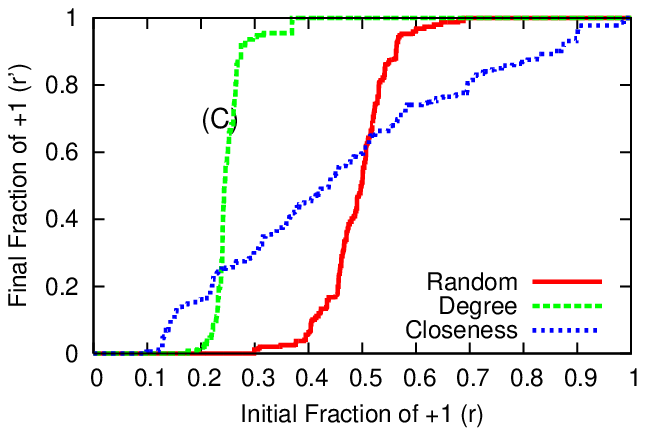}}  \\
   \multicolumn{3}{c}{ KE network } \\

   \resizebox{0.32\linewidth}{!}{ \includegraphics[]{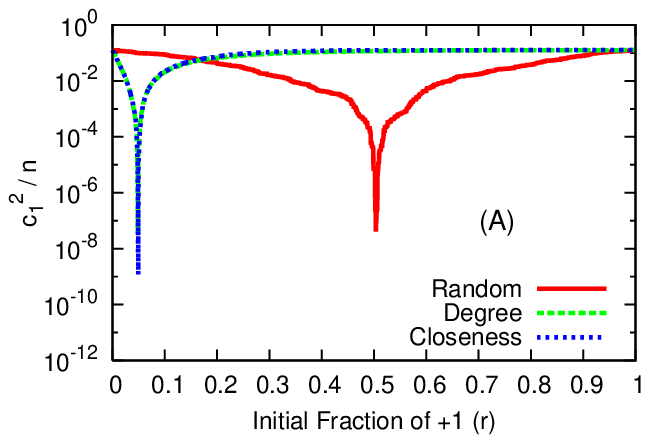}}  &
   \resizebox{0.32\linewidth}{!}{ \includegraphics[]{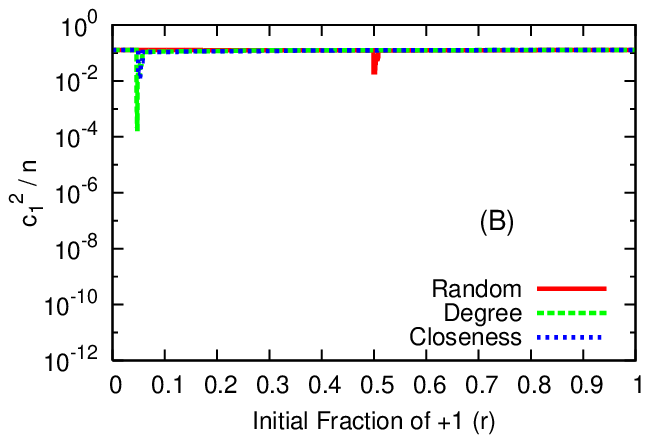}}  &
   \resizebox{0.32\linewidth}{!}{ \includegraphics[]{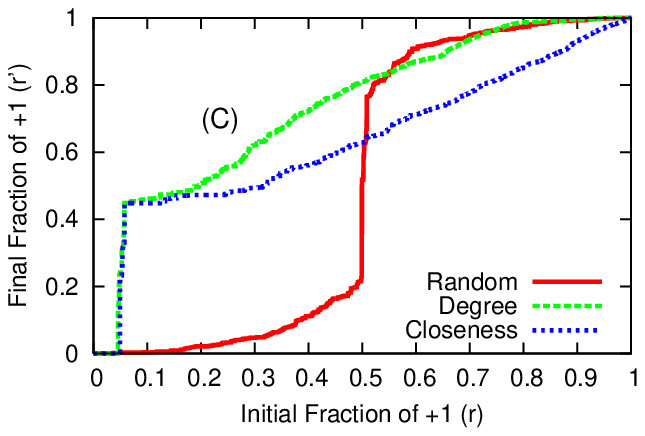}}  \\
   \multicolumn{3}{c}{ CNN network } \\

 \end{tabular}
  \caption{ Numerical results of 
  (A) Normalized magnitudes of the eigenmode of the largest eigenvectors on $\bm{s}_{0}$,
  (B) Normalized magnitudes of the eigenmode of the largest eigenvectors on $\bm{s}_{\infty{}}$, and
  (C) Fraction of $+1$ in $\bm{s}_{\infty{}}$, as function of $r$.  }
  \label{fig:init_random}
\end{center}
\end{figure}

First, we consider the initial state given by a random distribution.
%
%
The square of the first coefficient in the eigenmode 
expansion of $\bm{s}$ normalized by $n$ ($c_{1}^{2} / n$), 
which is derived from the largest eigenvector and minimizes the Hamiltonian
is examined. The solid lines plotted in Fig.~\ref{fig:init_random}(A) and (B) 
present $c_{1}^{2} / n$ versus the initial fraction of $+1$ for $\bm{s}_{0}$ 
and $\bm{s}_{\infty{}}$, respectively, 
where $\bm{s}_{\infty{}}$ denotes the final state.
%
%
Fig.~\ref{fig:init_random}(C) shows the fraction of $+1$ in 
the final state.
%
In the initial state, the largest eigenmode is prominent, except 
in the region of $r \simeq{} 0.5$. 
In that region, $c_{1}^{2} / n$ becomes close to zero. 
We have checked that the eigenmodes associated with smaller eigenvectors 
are dominant. Note that $\sum_{i}{c_{i}^{2} / n} = 1$. 
%
%
We also find the differences of the magnitude of $c_{1}^{2} / n$ among 
the networks.
In the ER, WS and BA networks, it is close to one, while it is about 0.1 
for the KE and CNN networks.
%
%
%
%
%
%
At $t = \infty{}$, $c_{1}^{2} / n$ is superior to the other 
associated with smaller eigenvectors in most range of $r$, 
as shown in Fig.~\ref{fig:init_random}(B).
The distributions of magnitude of $c_{1}^{2} / n$ among the networks 
are different. In the ER and the WS networks, $c_{1}^{2} / n$ is close to one. 
%
%
We have checked that the magnitude of $c_{2}^{2} / n$ for 
the KE network is almost same order of the $c_{1}^{2} / n$.
%
%
%
%
%
From Fig.~\ref{fig:init_random}(B) and (C),
relationships between $r - c_{1}^{2} / n$ and $r - r'$ relations 
can be exploited according to the change caused by the variation of $r$.
In the range of $r$ where the largest eigenvector is dominant, 
all vertices are in the stable state, 
that is, they have a single opinion.
%
%
The range that the two opinions coexist in the final state 
can be found from $r - r'$ relations. In such range of $r$, 
$c_{1}^{2} / n$ becomes smaller than the magnitude at the other range.

Secondly, we examine the initial state distribution determined 
by degree centrality and closeness centrality, whose results
are also represented in Fig.~\ref{fig:init_random}.
%
%
%
%
%
The range of $r$ where $c_{1}^{2} / n$ becomes smaller varies 
according to the initial conditions and network structures.
Let $r_c$ be the value of $r$ which $c_{1}^{2}/n$ takes nearly zero.
In the case of the ER and WS networks, $r_c$ is slightly less than $0.5$.
The value of $r_c$ for the other networks is much less than $0.5$. 
%
%
In $\bm{s}_{\infty{}}$, $c_{1}^{2} / n$ is dominant in most range of $r$,
while the value of $c_{1}^{2} / n$ is relatively small compared to the 
ER and WS networks.
%
%
The range of $r$ where $c_{1}^{2} / n \simeq{} 0$ in
$\bm{s}_{\infty{}}$ agree with that in $\bm{s}_{0}$.
%
%
Moreover, $r'$ in $\bm{s}_{\infty{}}$ clearly describe the characteristics: 
Around the value of $r_c$ in $\bm{s}_{\infty{}}$,
$r'$ rapidly increases as $r$.
Furthermore, in the range of $r$ where the dynamics
ends up in a metastable state with coexisting two opinions,
$c_{1}^{2}  / n$ has much smaller value,
 and it is assumed that the others associated with smaller eigenvectors 
 have relatively large value.


\section{ Discussion and Conclusion}

The eigenmode of the largest eigenvector implies 
the mode that all vertices have the same opinion. 
This corresponds to the initial state that 
the fraction of two opinions $r$ is $r=0$ or $r=1$
in $\bm{s}_{0}$.
On the other hand, one can consider a particular $r$
where $\bm{s}_{0}$ and the largest eigenvector $\bm{v}_{1}$
are almost orthogonal. 
Since the components of $\bm{v}_{1}$ are non-negative and
the components of $\bm{s}_{0}$ is $\pm{} 1$, 
$\bm{s}_{0}$ and $\bm{v}_{1}$ becomes almost orthogonal 
if randomly selected half of the components of $\bm{s}_{0}$
are $+1$, while the others are $-1$.
This is because the eigenmode of the largest eigenvector $c_{1}^{2} / n$
becomes $c_{1}^{2} / n \simeq{} 0$ in $\bm{s}_{0}$ at $r \simeq{} 0.5$,
if the initial state is randomly distributed.

Contrary to this, if the components of $\bm{s}_{0}$ whose corresponding
components of $\bm{v}_{1}$ are large, are preferentially assigned $+1$
at the initial state, the inner product $c_{1}$ of $\bm{s}_{0}$ and
$\bm{v}_{1}$ becomes $c_{1} \simeq{} 0$ although $+1$ is less 
than half in the components of $\bm{s}_{0}$, that is, $r < 0.5$. 
The value of $r$ is less than $0.5$ when the initial states are
distributed according to the centralities.  This suggests that
a certain correlation exists between the values of the elements of
$\bm{v}_{1}$ and such centralities. 
%
%
%
In the final state $\bm{s}_{\infty{}}$ the fractions of 
two opinions $r'$ steeply increase around the values of 
$r$ where
$c_{1}^{2} / n \simeq{} 0$ in $\bm{s}_{0}$. 
%
Such $r$ agrees with the value where $\bm{s}_{0}$
and $\bm{v}_{1}$ are orthogonal.
This fact result suggests that the final states 
and the values of $r$ at which $r'$ have rapid transition 
can be estimated by analyzing the magnitudes of 
the largest eigenmode on $\bm{s}_{0}$.


In summary, we have performed an eigenmode analysis on 
dynamics of the decision-by-majority process on complex networks.
Relationships between initial conditions and final states of 
the dynamics have been analyzed by eigenvectors of
adjacency matrices.
First, we confirmed analytically that the stable state in which
all vertices have the same opinion corresponds to the first
eigenmode of the system. 
Then, we analyzed the relationships between the dynamics 
induced by arbitrary initial conditions and eigenmodes
of networks computationally.
It has been shown that, the values of $r$
at which the magnitude of the largest eigenvector 
becomes extinct in the initial state,
gives $r'$ a rapid transition in the final state.
%
If the magnitude of the second or less eigenvector is
stronger than the largest eigenvector in the initial state,
a metastable state arise with two opinions coexisting
at the final state.
From this fact, the final state of the dynamics 
can be estimated by the magnitudes of the largest eigenmode
at the initial state.
In conclusion, we have thus shown that the eigenmode
analysis gives us a clue for understanding and predicting
an evolution of dynamics on complex networks.

%



\printindex
\end{document}